\documentclass[%
reprint,
superscriptaddress,
amsmath,amssymb,
prb
floatfix,
showpacs,
]{revtex4-2}

\usepackage{subcaption}
\usepackage{graphicx}
\usepackage{dcolumn}
\usepackage{bm}
\usepackage{hyperref}
\usepackage{ulem}

\usepackage[english]{babel}
\usepackage[utf8]{inputenc}
\usepackage{amssymb,amsfonts,amsmath,mathtools,mathrsfs}
\usepackage{relsize}
\usepackage{mdframed}
\usepackage[margin=0.6in]{geometry}
\usepackage{enumitem}
\usepackage{graphicx}
\usepackage{float}
\usepackage{url,hyperref}
\usepackage[table,usenames,dvipsnames]{xcolor}
\usepackage{gensymb}
\usepackage{grffile}

\hypersetup{
	colorlinks=true,
	linkcolor=Blue,
	filecolor=magenta,      
	urlcolor=Blue,
	citecolor=Blue,
}
\usepackage[bottom]{footmisc}
\usepackage[capitalize]{cleveref}
\usepackage{textcomp}
\usepackage{braket} 

\setcitestyle{line}

\usepackage{amsmath,amssymb,wasysym}

\makeatletter
\def\@fnsymbol#1{\ensuremath{\ifcase#1\or *\or \dagger\or \ddagger\or
   \mathsection\or \mathparagraph\or \|\or **\or \dagger\dagger
   \or \ddagger\ddagger \else\@ctrerr\fi}}
    \makeatother

\begin{document}

\title{Investigation of magnetic and magneto-transport properties in non-centrosymmetric antiferromagnetic semimetal GdGaSi}

\author{Manikantha Panda}
\address{Department of Physics, National Institute of Technology Andhra Pradesh, Tadepalligudem 534101, India}

\author{Prabuddha Kant Mishra}
\address{Department of Chemistry, Indian Institute of Technology Delhi, New Delhi 110016, India}
\address{Current Affiliation: Institute of Low Temperature and Structure Research, Polish Academy of Sciences, Okólna 2, 50-422 Wrocław, Poland}
\author{Sonali S Pradhan}
\address{Department of Physics, Indian Institute of Technology Hyderabad, Kandi - 502285, Sangareddy, Telangana, India.}
\author{V. Kanchana} \email[E-mail: ]
{kanchana@phy.iith.ac.in}
\address{Department of Physics, Indian Institute of Technology Hyderabad, Kandi - 502285, Sangareddy, Telangana, India.}

\author{Tapas Paramanik} \email[E-mail: ]
{tapas.phys@nitandhra.ac.in}
\address{Department of Physics, National Institute of Technology Andhra Pradesh, Tadepalligudem 534101, India}

\begin{abstract}
In this work, we investigated the magneto-transport and magnetic properties of GdGaSi, having non-centrosymmetric tetragonal structure, with space group $I4_1md$. Our theoretical results are supported by experimental studies. 
First-principles calculations reveal that GdGaSi is an antiferromagnetic semimetallic system, characterized by dominant electron-type charge carriers. In addition, the possible nontriviality of the crossing at the Fermi energy is consistent with isostructural LaPtSi-structured materials.
The compound shows robust antiferromagnetic (AFM) ordering with a Néel temperature of 19 K, and spin-reorientation signature below $T_N$. The semimetallic nature with positive magnetoresistance ($\simeq$ 2\% at 2 K and 8 T) is observed from the magnetotransport data, having electrons as majority charge carrier, established from the Hall measurements. The strong correlation in magnetism and transport is supported by various observations, like (1) concordant transitions in $M(T)$ and $\rho (T)$ data, (2) change in the concentration and mobility of electron below $T_N$, and (3) splitting of Kohler's plots in the two branches across the transition. 
 Thus, our findings establish GdGaSi as a material with intertwined magnetic and transport degrees of freedom, within noncentrosymmetric lattice.


\end{abstract}



\maketitle

\section{Introduction}

Correlated electron systems continue to captivate condensed matter physicists due to their ability to host a wide array of emergent quantum phenomena, including heavy fermion behaviour, unconventional superconductivity, huge magnetoresistnace and complex magnetic ordering. These properties often arise from the delicate interplay between itinerant conduction electrons and localized 4f-electron states in rare-earth elements, leading to a rich landscape of physical behaviours \cite{Steglich1979,mathur1998,Mishra2024b,Mishra2024c,Alam_CeAlSi,gaudet2021weyl,banik2018huge}.
Among such systems, ternary rare-earth intermetallic compounds, particularly those belonging to the $R$TX family (R = rare-earth elements such as Ce–Nd, T = Al or Ga, and X = Si or Ge), have recently emerged as promising candidates for the exploration of intertwined magnetism and topology \cite{MishraNdBiTe, Gautam2025}. These compounds display diverse magnetic characteristics depending on the choice of rare-earth ion. For instance, re-entrant spin glass behaviour has been reported in PrAlSi, while CeAlSi shows a non-collinear magnetic structure. Other members, like GdAlSi, exhibit field-induced antiferromagnetism, and compounds such as RGaSi and RGaGe display phenomena including anisotropic magnetocaloric effects and strong magnetocrystalline anisotropy \cite{Lou_PrAlSi2023, Gong_2024, Yang2021, ZHONG2025172855, Ram2023108, Zhang_2024cegasi}.
Importantly, several $R$TX compounds crystallize in two structural variants—orthorhombic (Cmcm, AlB$_2$ type) and non-centrosymmetric tetragonal ($I4_1md$, $\alpha$-ThSi$_2$ type) \cite{Wang_2020,WANG2021167739}. The latter structure has drawn increased attention due to its symmetry-breaking nature, which is a prerequisite for realizing exotic topological states like Weyl semimetals. Notably, the stabilization of these structures is sensitive to the Al/Si(Ge) ratio, which influences the resulting crystal symmetry \cite{DHAR199622}.  Indeed, a number of these $R$TX compounds have been identified as potential hosts of Weyl fermions, where broken time-reversal symmetry due to magnetic ordering leads to topologically protected features such as Weyl nodes, chiral anomaly, and the topological Hall effect \cite{gaudet2021weyl,Gong_CeGaSi,Ram2023108}.
Recent studies on Gd-based $R$TX systems, such as GdGaGe, have revealed complex magnetic transitions, including spin-flop behaviour and unusual magnetoresistance characteristics that reflect the intricate spin scattering mechanisms and anisotropic interactions \cite{cao2025}.

Atomic site disorder between p-block elements, like Ga/Al and Ge/Si, has been demonstrated to have a major impact on the structural symmetry in the $R$TX systems.  A centrosymmetric structure [space group $I4_1/amd$, (No.~141)] is adopted by the system in the presence of disorder, while a noncentrosymmetric LaPtSi-type structure [space group $I4_1md$, (No.~109)] is produced by an ordered arrangement.  The possible topological characteristics of the material are directly impacted by this structural difference \cite{ZHONG2025172855,Puphal24204}. In this context, GdGaSi stands out as a relatively unexplored member of the $R$TX family. The substantial size mismatch between Ga and Si (r$_{Ga}$/r$_{Si}$) $\sim$ 1.80) reduces the likelihood of inter site mixing compared to other $R$TX compounds such as $R$AlSi and $R$GaGe, thereby preserving the non-centrosymmetric tetragonal structure critical for topological properties \cite{Jadupati}. Furthermore, the use of Gd, with its half-filled 4$f^7$ shell and zero orbital angular momentum (L = 0), minimizes crystal field effects while providing a strong localized magnetic moment, making it an ideal probe for studying the interplay between magnetism and electronic structure. Although an initial report on GdGaSi exists \cite{weitzer2013physical}, comprehensive insight into its magnetic and magneto-transport behaviour remains lacking. Given the growing interest in topologically nontrivial and altermagnetic systems, a thorough investigation of GdGaSi is both timely and significant.
In this study, we present a detailed exploration of the structural, magnetic, and electronic properties of GdGaSi. Through a combination of experimental techniques and band structure analysis, we aim to establish its magnetic ground state, assess its magneto-transport behaviour, and evaluate its potential as a platform for correlated topological magnetism. Our findings provide new insight into how electronic transport, crystal symmetry, and magnetism interact in rare-earth-based intermetallics.

\section{METHODOLOGY}
First-principles calculations were performed using the projector augmented-wave (PAW) method \cite{blochl1994projector}, as implemented in the Vienna \textit{ab initio} simulation package (VASP) \cite{kresse1996efficiency,kresse1999ultrasoft}, based on density functional theory (DFT) \cite{hohenberg1964inhomogeneous}. The Perdew--Burke--Ernzerhof (PBE) \cite{perdew1996generalized} exchange-correlation functional was used, and the convergence threshold of the Hellmann--Feynman force was set to 0.01~eV/\AA. A plane-wave energy cutoff of 600~eV was consistently applied in all calculations, with the electronic energy convergence criterion set to $10^{-8}$~eV. Geometry optimizations were performed using a dense $\Gamma$-centered $12 \times 12 \times 6$ $k$-point mesh to sample the irreducible Brillouin zone, following the Monkhorst-Pack scheme \cite{monkhorst1976special}. To account for the strong correlation effects of the Gd \textit{f}-states, the GGA+$U$ approach was employed with an effective Hubbard $U$ parameter of 6~eV, following the method proposed by Dudarev \cite{dudarev1998electron}. For studying the experimental findings polycrystalline sample of GdGaSi was prepared by the conventional arc melting technique under a high-pure Ar gas atmosphere. As starting materials, stoichiometric amounts of constituent elements of high purity ($> 99.9\%$) were taken in the form of pieces. To achieve better homogeneity, the ingot was remelted five times by flipping it on both sides. The as-cast ingot was vacuum sealed in a quartz tube and annealed at 900 $^{\circ}$C for 10 days in a tubular furnace. A negligible weight loss ($< 0.50\%$) was observed for the resulting sample. For magneto-transport measurement, a small piece of rectangular parallelepiped shape of dimension 3.84$\times$1.9$\times$0.7 $mm^3$ (l$\times$w$\times$h) was cut from the interior of the annealed ingot. The Rietveld refinement of room-temperature PXRD data using TOPAS software package \cite{Topas}, confirmed single-phase formation and was used for the determination of the crystal structure of the sample. With a FESEM (TESCAN MAGNA), compositional analysis was carried out. Using a 5 T magnet and an MPMS3 SQUID magnetometer (Quantum Design), temperature and field dependent magnetization measurements were made.  In a Quantum Design PPMS system, resistivity as a function of temperature (2-300 K) and magnetic field (8 T) was measured using the conventional four-probe approach. 


\section{THEORETICAL RESULTS}
\subsection{Crystal Structure and Magnetic Properties}

\begin{figure}[htb]
\includegraphics[width=1\columnwidth]{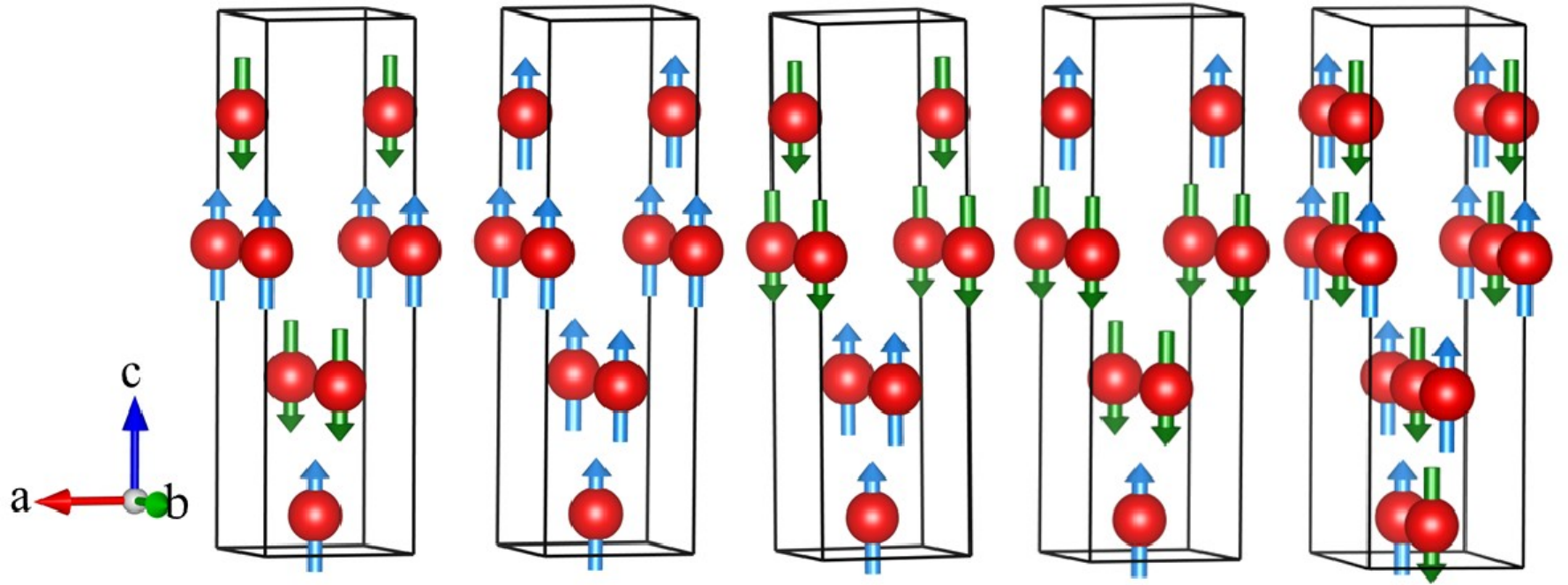}
        \caption{Magnetic configurations of GdGaSi with Gd spins:  AFM1, FM, AFM2, AFM3, and AFM4 (In-plane AFM), respectively. Spin-up and spin-down states are represented by cyan and green arrows, respectively.}
        \label{conf}
\end{figure} 

\begin{figure*}
    \centering
    \includegraphics[width=1\linewidth]{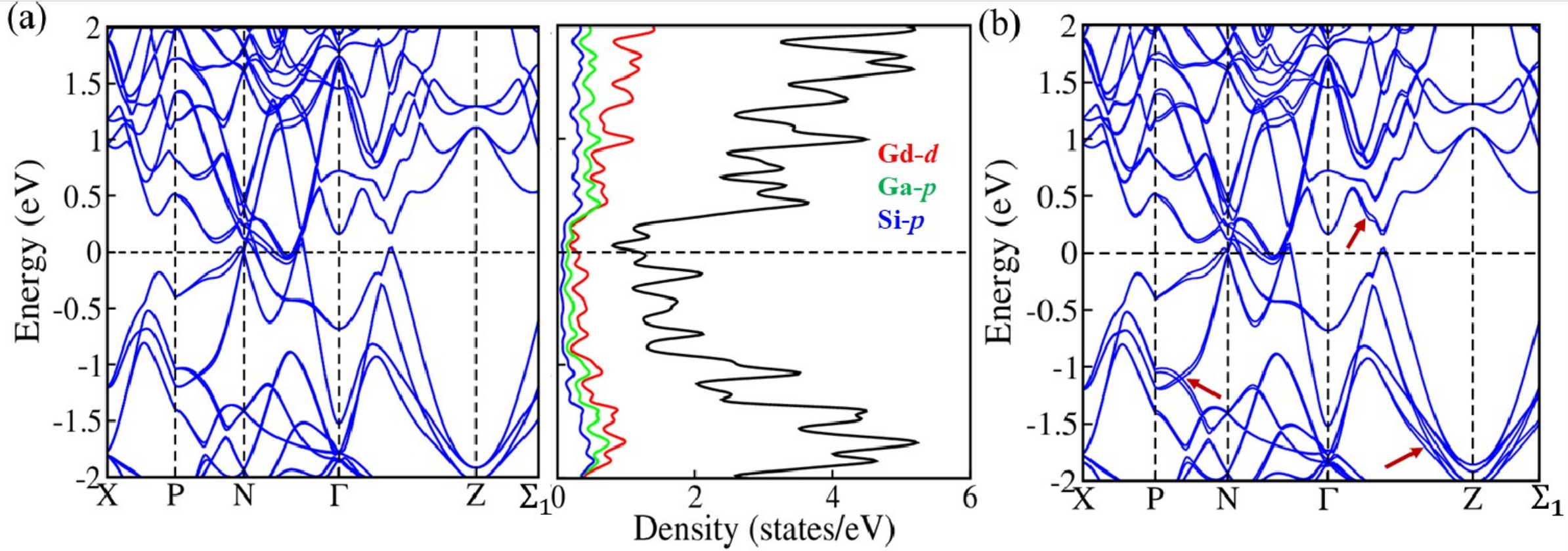}
    \caption{(a) Electronic band structure without SOC, along with the total and projected density of states. (b) Electronic band structure with SOC.}
    \label{ba}
\end{figure*}
In \textit{R}GaSi compounds, Ga/Si site disorder is commonly observed, where gallium and silicon atoms randomly occupy each other’s Wyckoff positions. This disorder typically leads to a centrosymmetric with space group \textit{I}4$_1$/\textit{amd} (No.~141), as previously reported for compounds in a related family~\cite{Jadupati,Puphal24204}. In contrast, when Ga and Si retain their distinct and well-defined Wyckoff sites without intermixing, the structure adopts a non-centrosymmetric space group \textit{I}4$_1$\textit{md} (No.~109), as illustrated in Fig.~4(b). To verify this, we performed first-principles total energy calculations for both the ordered and disordered structures. The results confirm that the compound stabilizes in the noncentrosymmetric \textit{I}4$_1$\textit{md} (No.~109) configuration, which is 2.046~eV lower in energy than the Ga/Si-disordered (centrosymmetric) structure. Therefore, we proceed with the lower-energy noncentrosymmetric ground-state structure in our analysis.\\

From the first-principle calculations, the lattice parameters are found to be $a = b = 4.14$~Å and $c = 14.23$~Å for LaPtSi-type structure of GdGaSi.  
The magnetic ground state was determined by comparing different magnetic configurations, as illustrated in \hyperref[conf]{Fig. 1}. Table \ref{tab} summarizes the corresponding energy differences between these magnetic states. Our calculations reveal that all antiferromagnetic (AFM) configurations are energetically more favorable than the ferromagnetic (FM) phase, with significant differences in their relative energies. In particular, the AFM1 configuration, characterized by in-plane ferromagnetic and out-of-plane antiferromagnetic spin alignment (udud), is found to be the lowest in energy.

\begin{table}[h]
\centering
\caption{Calculated  energies of different magnetic configurations (in meV), with the reference energy set to 0 meV.}
\setlength{\tabcolsep}{1.5pt} 
\renewcommand{\arraystretch}{1.22} 
\begin{tabular}{@{}c *{5}{>{\centering\arraybackslash}p{1.1cm}}@{}} 
\hline\hline
Configuration & AFM1 & FM & AFM2 & AFM3 &  AFM4 \\ 
\hline
Energy/f.u. (meV) & 0 & 57.1 & 29.56 & 29.65 & 20 \\
\hline
\label{tab}
\end{tabular}
\end{table}

To determine the preferred spin orientation in the AFM1 phase, we computed the ground state energies for several spin directions, including [001], [010], [100], [011], [101], [110], and [111]. Among these, the [010] orientation yields the lowest energy. The estimated effective magnetic moment , $\mu_{\mathrm{eff}} = 7.55\ \mu_B$, closely matches the theoretical value of $7.12\ \mu_B$/Gd expected for the Gd$^{3+}$ ion.\\

\subsection{Electronic properties}
We calculated the electronic band structure, as shown in \hyperref[ba]{Fig.~2(a)}, which reveals a slight overlap between the valence and conduction bands near the Fermi level, indicating semimetallic behaviour. The corresponding total density of states (DOS), along with the atom-resolved projected density of states (PDOS), is also presented in \hyperref[ba]{Fig.~2(a)}, highlighting significant contributions from the Gd-$d$, Ga-$p$, and Si-$p$ orbitals. The valence band is primarily dominated by Gd states, with Ga and Si contributing almost equally, while the conduction band is predominantly governed by Gd states. The reduced DOS at the Fermi level further confirms the semimetallic nature of the system. Upon applying an appropriate DFT+$U$ correction, the Gd-$f$ electrons become highly localized, resulting in negligible contributions to the electronic states near the Fermi level.

The incorporation of spin-orbit coupling (SOC) enables the evaluation of the magnetocrystalline anisotropy energy (MAE). When the magnetic moment lies within the $ab$ plane ($M \perp c$), the system exhibits a lower total energy compared to the configuration with $M \parallel c$. In our case, the MAE is relatively small, with a value of approximately 0.07~meV. Additionally, the energy difference between orientations of $M$ along the $x$ axis and the $y$ axis is negligible. These findings indicate that the $ab$ plane acts as the easy plane of magnetization. This kind of MAE along different crystallographic direction has already been observed in PrAlSi and CeGaSi \cite{Lyu2020,Gong2024_anisotropy} whereas  GdAlSi exhibits very weak MAE, with a susceptibility ratio $\chi_c/\chi_{ab}$ = 0.95 \cite{Laha2024109}. Moreover, SOC lifts band degeneracies and opens band gaps along various paths in the Brillouin zone, as indicated by the red arrows in \hyperref[ba]{Fig. 2(b)}. However, its effect on the electronic band structure near the Fermi level remains relatively weak.

\begin{figure}[h]
\begin{center}
        \includegraphics[width=1\columnwidth]{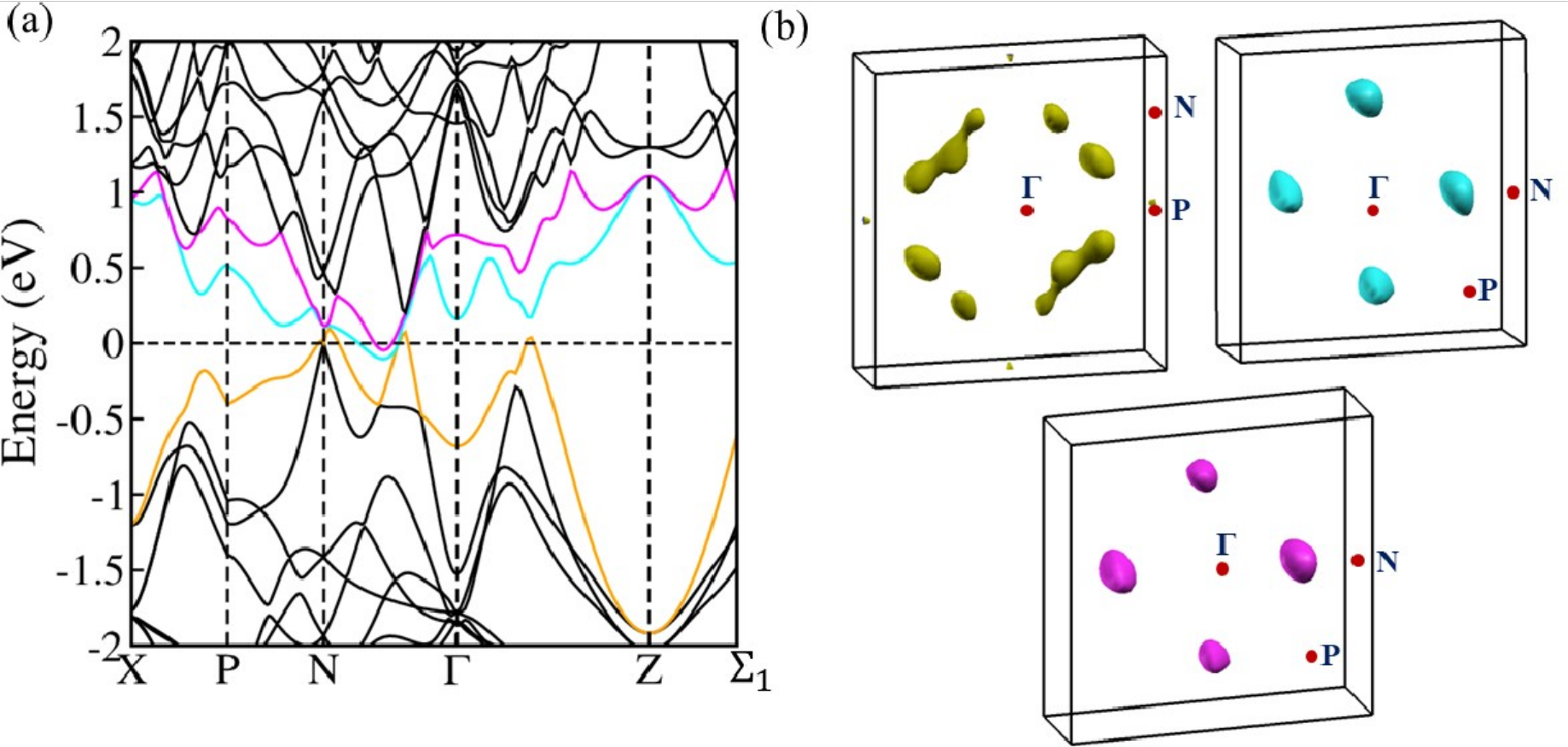}
        \caption{ (a) Electronic band structure without SOC, with colored bands intersecting the Fermi level.
(b) The first one is a hole-type, while the last two are electron-type Fermi surfaces.}
        \label{EF}
\end{center}
\end{figure}

The Fermi surface plots for the three bands crossing the Fermi level in the absence of SOC are shown in \hyperref[EF]{Fig. 3(b)}. The first Fermi surface, corresponding to the orange band, is of hole-type character and intersects the Fermi level from the valence band to the conduction band near the N, $\Gamma$, and Z points. In contrast, the remaining two Fermi surfaces, corresponding to the magenta and cyan bands, are of electron type and intersect the Fermi level from
the conduction band to the valence band in between the N and $\Gamma$ points. Overall, the contribution of electron-type carriers is greater than that of hole-type carriers for the compound. Furthermore, in order to confirm the consistency between theory and experiment, we conducted extensive experimental investigations on the compound GdGaSi, emphasizing on its structural, magnetic, and electronic properties.
\section{EXPERIMENTAL RESULTS}
\subsection{X-ray diffraction}
\begin{figure*}
\begin{center}
    \includegraphics[width=0.8\linewidth]{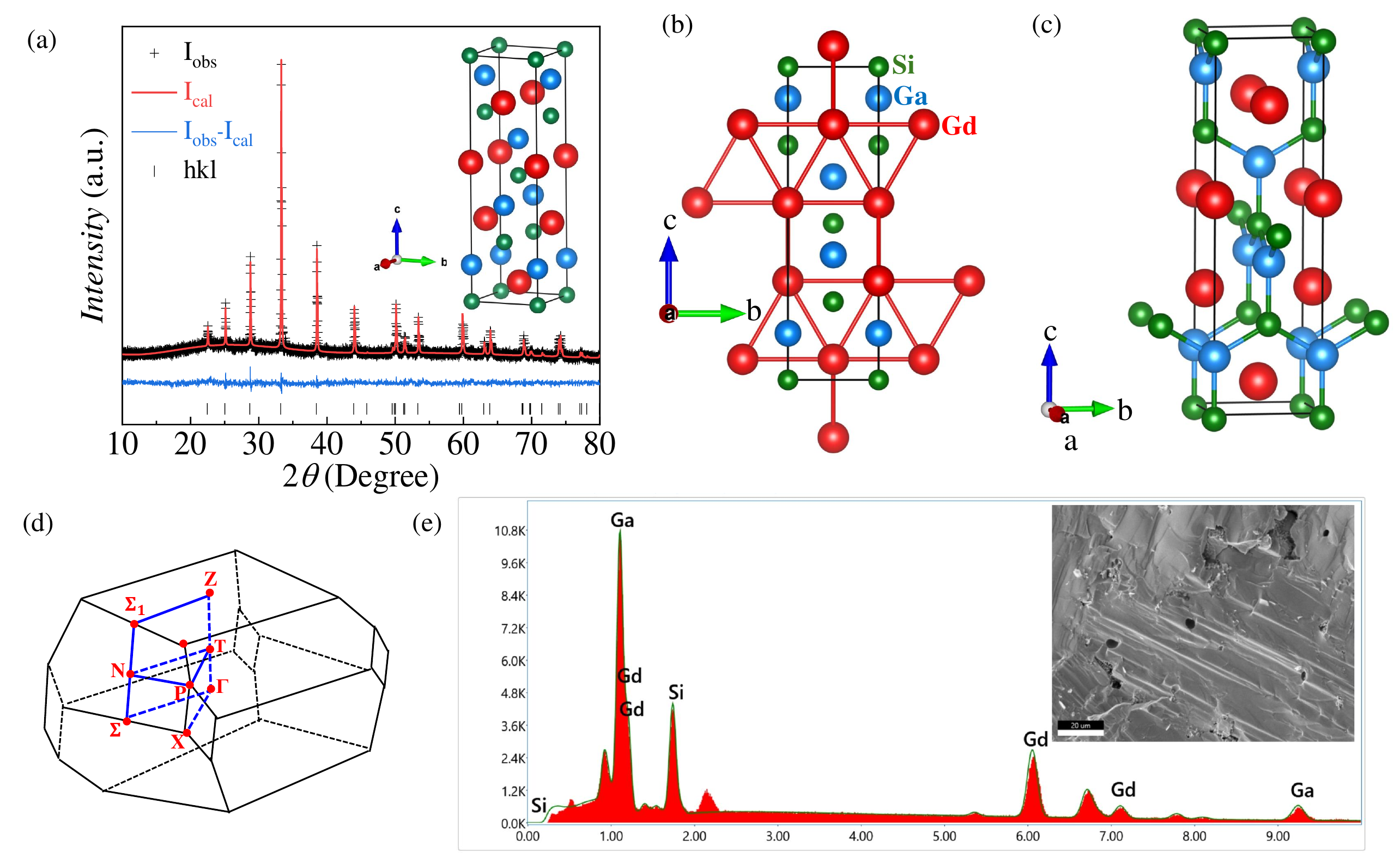}
    \caption{(a)The Rietveld refined PXRD of the GdGaSi polycrystal.  The difference between the measured and computed intensities is shown by the blue line.  Bragg's reflections that are allowed for GdGaSi are shown by the green vertical bars.(Inset shows the crystal structure of GdGaSi); (b) View along the a-axis of unit cell, with highlighted isosceles triangles of Gd, (c) Unit cell displaying Gd enclosed in Ga-Si framework, (d) Bulk irreducible Brillouin zone illustrating high-symmetry points and (e) Spectroscopy data for GdGaSi from EDX. The stacking nature of the sample is shown in the inset from FESEM image.}
    \label{fig1}
\end{center}
\end{figure*}


The powder X-ray diffarction (PXRD) pattern of finely crushed GdGaSi has been recorded at room temperature, as presented in \hyperref[fig1]{Fig. 4(a)}. 
The Rietveld refinement of PXRD data suggests that GdGaSi crystallizes in a noncentrosymmetric tetragonal structure with space group $I4_1md$ \cite{Weng2015,Huang2015}. The parameters obtained from the refinement are tabulated in the Table \ref{Table I}, are in excellent agreement with the theoretically obtained value. As shown in the \hyperref[fig1]{Fig. 4(c)}, the structure consists of four formula unit per unit cell, in which all three atoms (Gd, Ga, and Si) have 4a Wyckoff  position with same symmetry. There is stacking of Gd, Ga, and Si layers along the (001) direction. This structure has two vertical mirror planes ($\sigma_v$) but lacks a horizontal mirror plane ($\sigma_h$), resulting in inversion symmetry breaking.
The structure of GdGaSi can be visualized as alternatively stacked layers of Gd, Ga, and Si in the unit cell.

The EDAX data are taken on several areas of the specimen, yielding an average chemical composition of Gd:Ga:Si as 1.02:0.99:0.99. The obtained ratio is close to equiatomic stoichiometry, as expected from the formula.

\begin{figure}[t!]
\includegraphics[width=0.8 \columnwidth,angle=0,clip=true]{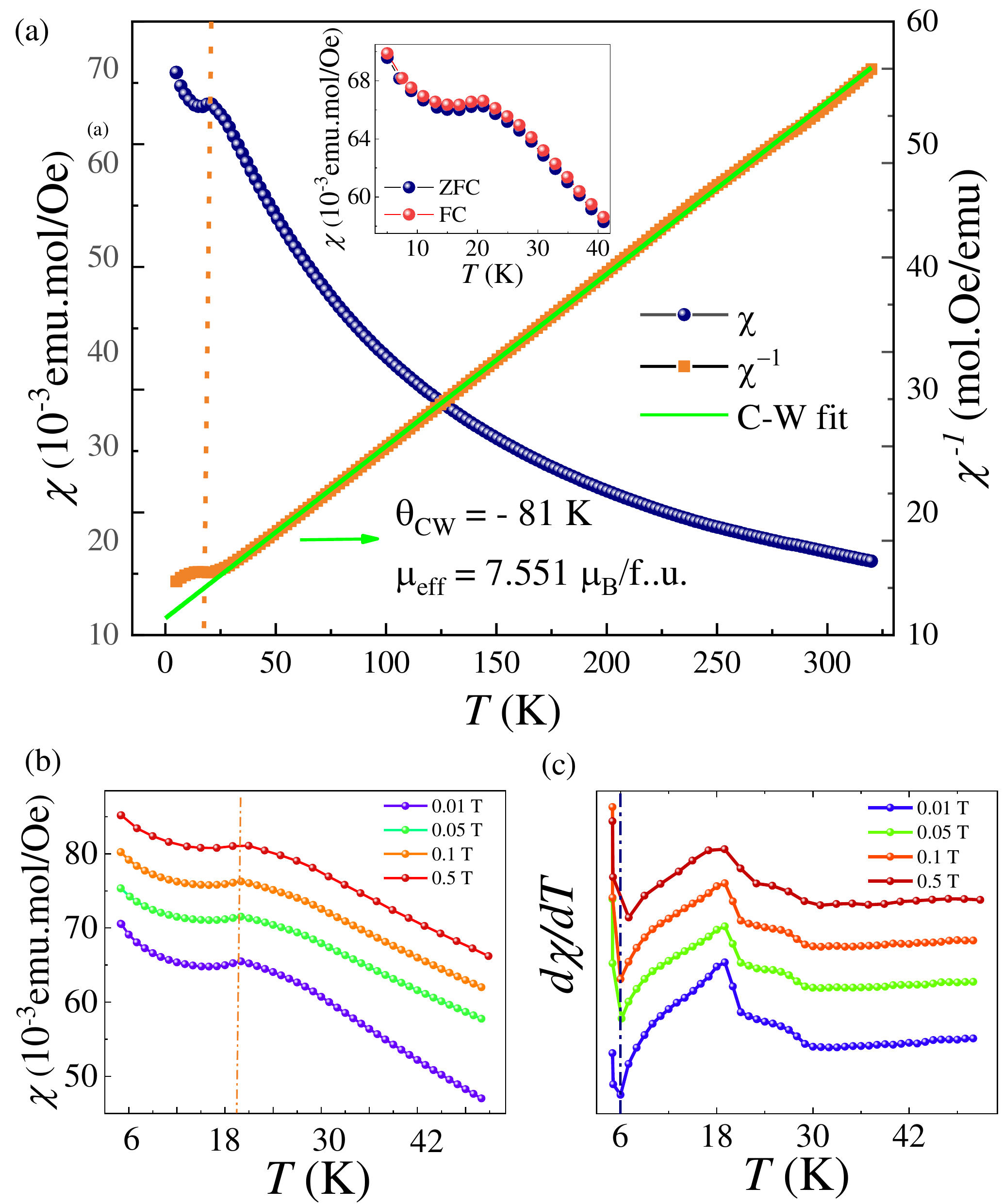}
\caption{(a) ZFC and FC modes temperature dependency of magnetic susceptibility ($\chi$) under 0.5 T applied field in the 2–320 K range (left axis).  Together with the Curie–Weiss fit (green solid line) carried out in the 25–320 K range (right axis), the inverse susceptibility ($\chi^{-1}$) for the FC data is also displayed.  An enhanced picture of the low-temperature $\chi$ vs. T curves under ZFC and FC procedures is displayed in the inset.
 (b) $\chi$ vs. T curves obtained at 5–50 K in a range of magnetic fields.
 (c) The derivative $d\chi/dT$'s temperature dependency under various applied magnetic fields.}
\label{fig2}
\end{figure}


\begin{figure}
\begin{center}
    \includegraphics[width=0.8\linewidth]{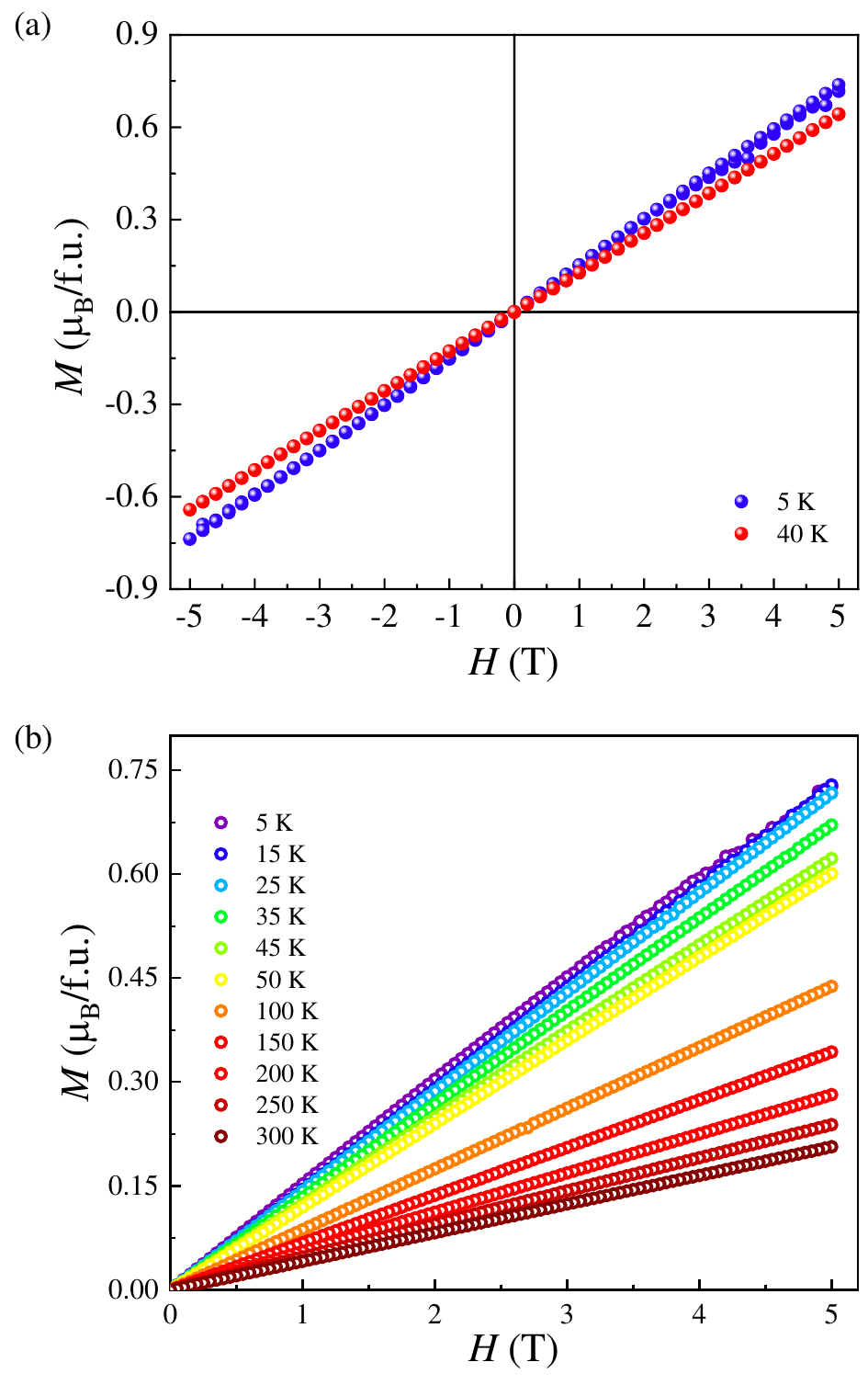}
    \caption{Field-dependent magnetization $M(H)$ (a) Isothermal magnetization taken at 5 and 40 K. (b) The virgin isotherms taken at different temperatures.}
    \label{fig3}
\end{center}
\end{figure}

\begin{table}[h]
\scriptsize\addtolength{\tabcolsep}{-1pt}
\caption{Structural and magnetic parameters of GdGaSi obtained from experimental measurements.} \label {Table I}
\centering
\begin{ruledtabular}
\begin{tabular}{l c c c c c c}
Space group: & $I4_1md$ \\
Space group number: & 109\\
{\it a}(\r{A}): & 4.12(2) \\
{\it c}(\r{A}): & 14.21(3) \\
\hline

Atom & Site & x & y & z & Occu. & B$_{iso}$\\
Gd & 4a & 0 & 0 & 0.574(3) & 1 & 0.78(9) \\
Ga & 4a & 0 & 0 &  0.157(1) & 1 & 0.76(9) \\
Si & 4a & 0 & 0 & 0 & 1 & 0.85(1) \\

\hline

$T_N$ &&&& \multicolumn{3}{r}{19 $\pm$ 0.5 K}\\
$\Theta_P$ &&&& \multicolumn{3}{r}{-81.1 $\pm$ 0.5 K}   \\
$C$  &&&& \multicolumn{3}{r}{7.1 $\pm$0.05 emuOe$^{-1}$mol$^{-1}$K$^{-1}$}   \\
$\mu_{eff}^{theo}$ &&&& \multicolumn{3}{r}{7.93$\mu_B$/Gd$^{3+}$}   \\
$\mu_{eff}^{calc}$ &&&& \multicolumn{3}{r}{7.55$\pm$0.01$\mu_B$/Gd$^{3+}$}   \\
$M_S$ at 5 K &&&& \multicolumn{3}{r}{0.78$\pm$0.02 $\mu_B$/ f.u}   \\

\end{tabular}
\end{ruledtabular}
\end{table}

\subsection{DC magnetization study}

We have measured magnetization of GdGaSi as a function of temperature and have investigated the details of the observed transition. The behaviour of DC magnetic susceptibility in the temperature range of 2-320 K, under an applied field of 0.5 T, is shown in \hyperref[fig2]{Fig.5(a)}. From the $\chi$ vs $T$ plot, it has been observed that there is a kink in $\chi$ at a temperature $T
_N\sim$ 19 K, which signifies the phase transition of moments from paramagnetic (PM) to the anti-ferromagnetic (AFM) ordered states, which aligns well with the theoretical result. This observed transition temperature agrees well with the value reported in the literature for this sample \cite{weitzer2013physical}. 
To examine the presence of other phase in the low temperature ($T<T_N$), $\chi$ vs $T$ measurement were done in different protocols. Usually, due to competing interactions, the magnetic irreversibility is reflected as bifurcation in the zero field cooled (ZFC) and field cooled (FC) data. To confirm it, we have performed measurements under both ZFC and FC protocols.
There is no irreversibility between ZFC and FC magnetization which infers AFM as dominant interaction among moments and rules out any coexisting phases below $T_N$. Additionally, the inverse of magnetic susceptibility ($\chi^{-1}$) plotted against temperature ($T$) from 5 K to 320 K has been analyzed for the purpose of quantifying magnetic susceptibility parameters. 
It can be seen from the \hyperref[fig2]{Fig. 5(a)} that the $\chi^{-1}$ varies linearly with temperature (for $T> 25 K$), which is commonly observed in Gd-based samples due to the least crystal field effect for symmetrically filled $f^7$ electronic states. On the right axis of \hyperref[fig2]{Fig. 5(a)}, the curve is fitted using the Curie–Weiss law.  

\begin{equation}\label{CW}
\chi (T) =  \frac{C}{(T-\theta_p)}
\end{equation}

The effective magnetic moment ($\mu_{eff}$) of the magnetic ions in the material can be inferred from the Curie constant ($C$) using the formula $C = N_A\mu_{eff}^2/3k_B$, where $N_A$ is Avogadro's number, $k_B$ is the Boltzmann constant, and $\theta_p$ is the Weiss temperature.  The Curie-Weiss fit (shown in solid green line) in the temperature range of 25 K to 320 K yielded an effective magnetic moment of 7.55 $\mu_B$/Gd$^{3+}$. This result is in agreement with the theoretical value for the Gd$^{3+}$ ion, which is $\mu_{theo}$= $[S(S+1)]^{1/2} \mu_B$ = 7.93 $\mu_B$.  $\theta_p$ has a calculated value of -81 K. where the negative sign implies that the ground state of the compound is dominated by AFM interactions among Gd$^{3+}$ spins. 

Generally, the magnetic susceptibility drops sharply below the AFM transition, due to complete compensation among antiferromagnetic change among neighbouring moments \cite{coey2010magnetism}. In our case, $\chi$ has a kink at $T_N$ and rises again as the temperature is lowered below the transition. This feature may originate from a spin-reorientation effect, which has also been observed in other Gd-based systems \cite{Gautam2025}. Notably, it persists under applied magnetic fields ranging from 0.01 T to 0.5 T and is more distinctly visible in the derivative plots of susceptibility with respect to temperature [see \hyperref[fig2]{Fig. 5(c)}].


$M(H)$ measurements were performed under applied magnetic fields up to ±5 T (shown in \hyperref[fig3]{Fig. 6}) in order to better investigate the nature of the magnetic states across the temperature range. AFM nature of the aligned moments is confirmed by the absence of hysteresis and the linear dependency of magnetization on the applied field.  There is no discernible departure from linearity across the Néel temperature.  The highest magnetization at 5 K is 0.78 $\mu_B$/f.u., which is almost 0.1 times the theoretical saturation magnetization ($M_s \sim$ 7.93 $\mu_B$/f.u.) anticipated for the Gd$^{3+}$ ion ($gJ\mu_B$).
The significantly smaller moment value suggests a robust AFM phase least affected by the field and has a negligibly small field-induced response. The observed features of linear magnetization, absence of field-induced response and magnetization values are close to sister compound GaAlSi \cite{Jadupati}.




\begin{figure}
\begin{center}
    \includegraphics[width=0.8\linewidth]{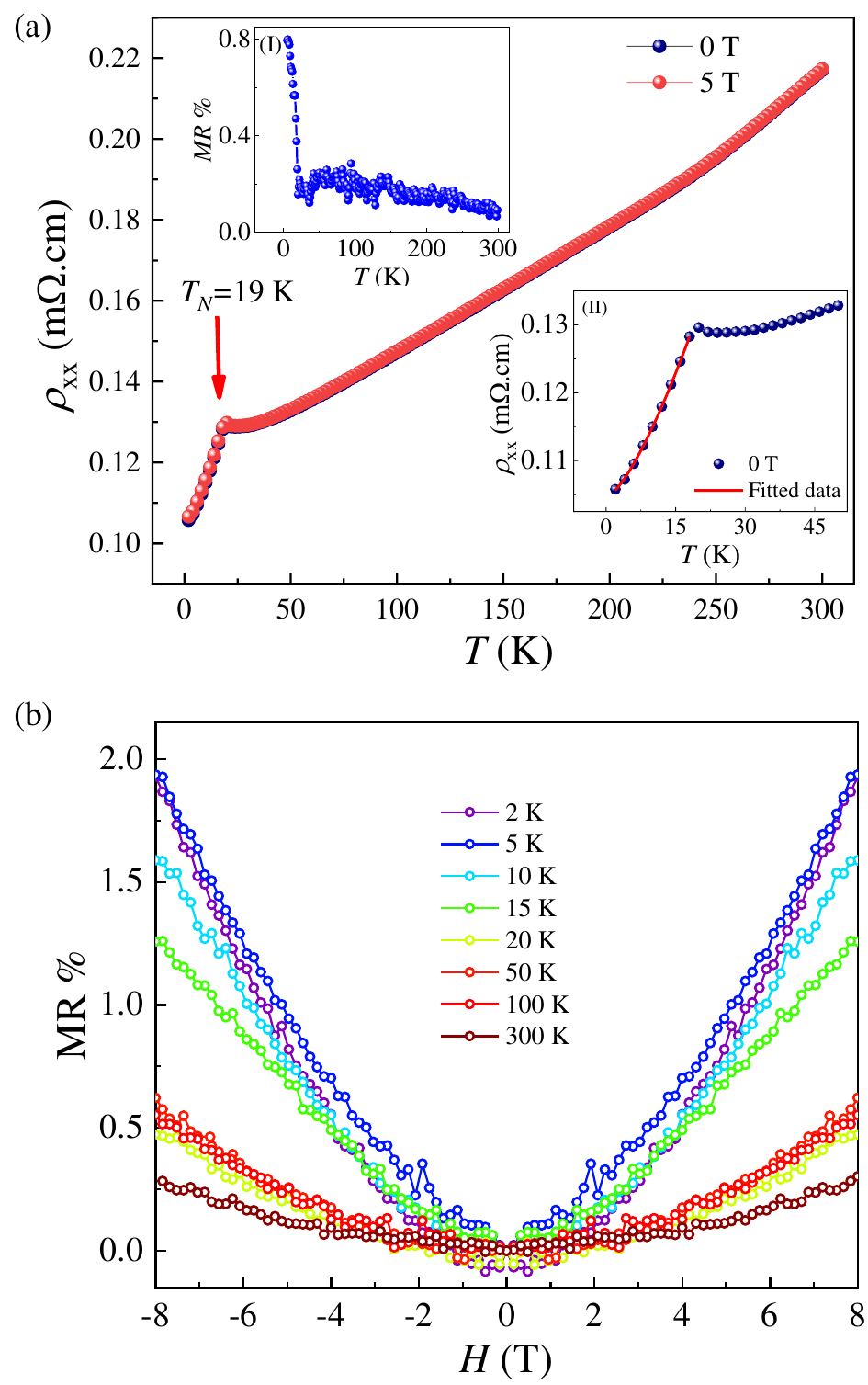}
    \caption{(a) Temperature-dependent longitudinal electrical resistivity of GdGaSi (H = 0 T and 5 T) in the temperature range of 2 K - 300 K. The inset (I) shows the MR as a function of temperature and inset (II) shows the fitting of resistivity data (2-19 K) to equation 3; (b) Field-dependent magnetoresistance under applied magnetic fields ranging from -8 T to +8 T, measured at different temperatures.}
    \label{fig4}
\end{center}
\end{figure}

\begin{figure}
    \centering
    \includegraphics[width=0.8\linewidth]{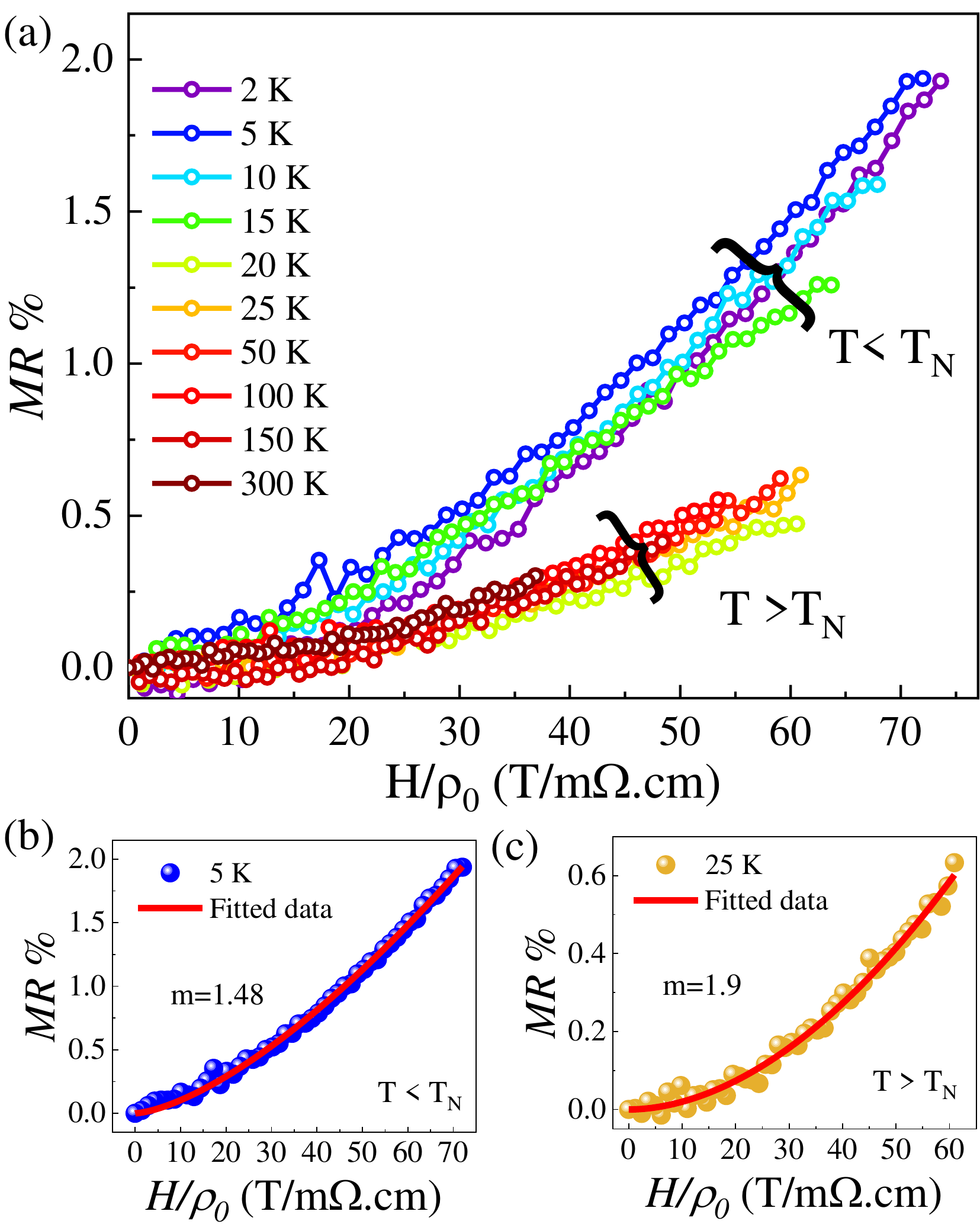}
    \caption{(a) Kohler's analysis is implemented to scale magnetoresistance at various temperatures, (b) and (c) Power law (Eq-4) fitting of the experimental data at 5 K and 25 K, respectively.}
    \label{fig5}
\end{figure}

\begin{figure}
    \centering
    \includegraphics[width=0.8\linewidth]{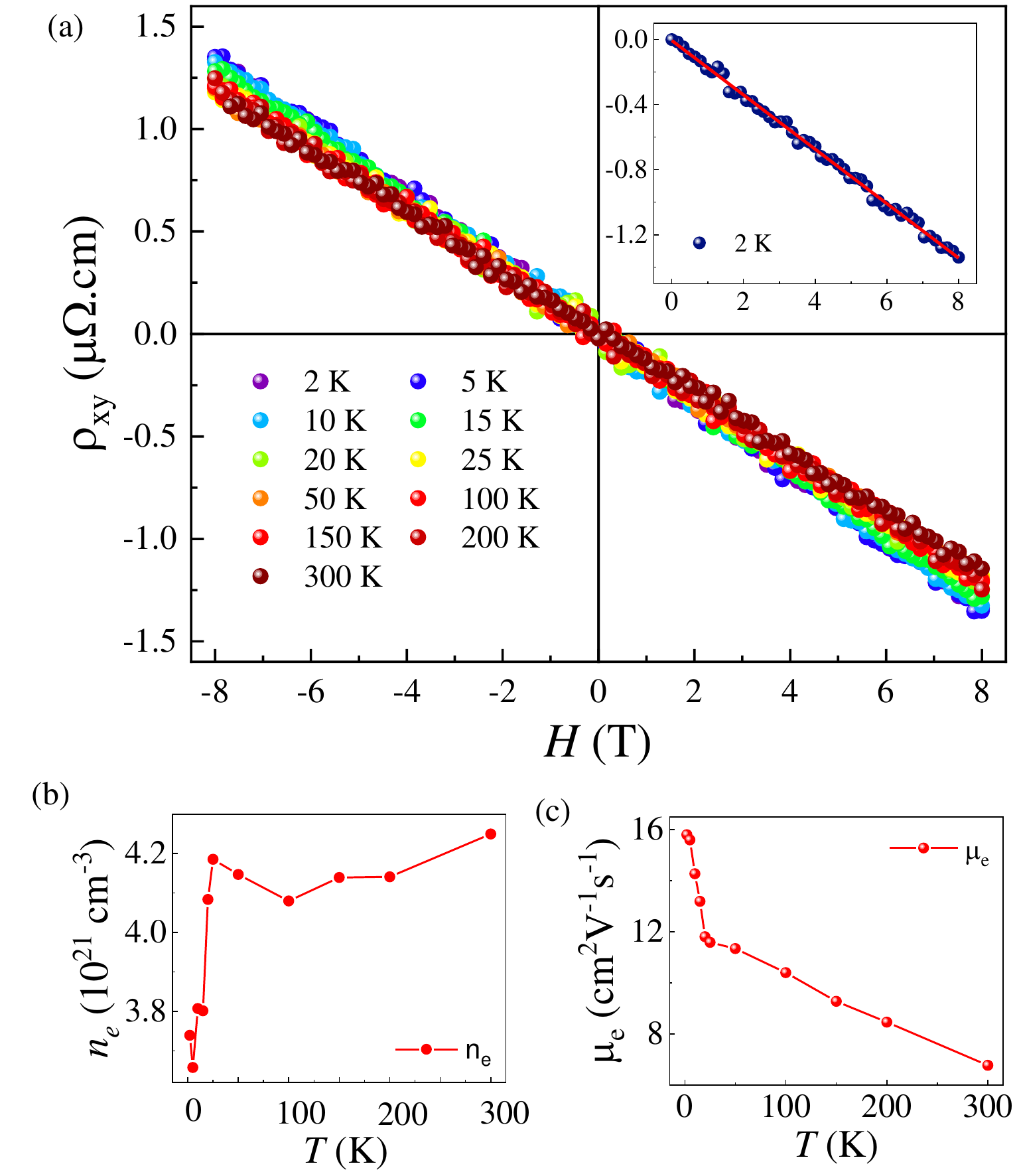}
    \caption{(a) Hall resistivity in the field range of $\pm$8 T at different temperatures ranging from 2 K to 300 K (inset shows the linear fit at 2 K using the single carrier model). (b) and (c) The temperature-dependent carrier density ($n_e$) and mobility ($\mu_e$) obtained from a single carrier model fit.}
    \label{fig6}
\end{figure}


 \subsection{Transport and magnetotransport studies}

A much interest have been grown in transport and magnetotransport properties of RGaX (R= rare earths, X= Ge, Si) materials, as they are candidate for nontrivial topology in electronic band structure \cite{ZHONG2025172855,Ram2023108}.
At first, we study the temperature dependence of resistivity to get insight into the relationship between magnetism and the transport characteristics of charge carriers in GdGaSi.  \hyperref[fig4]{Fig. 7(a)} illustrates the longitudinal resistivity $\rho(T)$, measured at zero magnetic field and under an applied field of $H = 5$ T.  The positive $d\rho/dT$ slope in both situations denotes semimetallic behaviour.
From the zero applied field data, we have calculated residual resistivity ratio (RRR=$\rho_{300 K}/\rho_{2K}$) and found it to be 2.06 which is comparable to it's isostructural compound NdAlGe \cite{Yang2021}. Such low RRR values are also found in the GdAlSi and related materials \cite{Jadupati, Gong_2024}. 
A clear signature associated with magnetic ordering has been observed in $\rho(T)$ data, the resistivity drops sharply below $T_N$, due to reduced scattering of carriers in ordered magnetic state. Similar behaviour of correlated characteristics in magnetic and transport properties has been observed in GdAlSi\cite{Laha2024109} 
The linear temperature dependence of $\rho(T)$ for temperature above the $T_N$ in GdGaSi suggests that the electron-phonon scattering is dominant.

The $\rho(T)$ data under applied field of 5 T, has same temperature dependence and have values in the similar range. Prima facie, it seems overlap the zero field $\rho(T)$ data. A more informative picture can be obtained when we consider magnetoresistance as function of temperature, as shown in the inset of \hyperref[fig4]{Fig. 7(a)}. MR is computed using the formula $\frac{\rho(H)-\rho(0)}{\rho(0)} \times 100 \%$, where $\rho(H)$ and $\rho(0)$ represent the resistivity with and without the applied magnetic field, respectively.  It is clear that MR shows a noticeable increase below the magnetic transition temperature and stays positive throughout the whole temperature range of observation (2–300 K).

The temperature-dependent resistivity $\rho(T)$ can be disentangled into various contributions as written below;

\begin{equation}\label{resistivity}
    \rho(T)=\rho_0+\rho_{ee}+\rho_{ep}
\end{equation}

where $\rho_0$, $\rho_{em}$ and $\rho_{ep}$ are the residual resistivity, electron-electron scattering coefficient and elctron-phonon scattering coefficient, respectively. $\rho_0$ is independent of both temperature and magnetic ordering, and primarily originated though lattice defects. In the high temperature regime, the resistivity is dominated by the third term $\rho_{ep}$, and is scaled linearly with $T$. 
Below the transition ($T<T_N$), $\rho(T)$ can be fitted by the following equation;

\begin{equation}
    \rho=\rho_0+aT^n
\end{equation}

where the second term is due to the electron-electron (e-e) scattering of conduction electrons, a  denote the e-e scattering coefficient and n is the exponent which varies depending upon the scattering mechanism. Usually, where the e-e scattering is dominated, the n value gives rise to be 2. However, the best fit to our data yields the n value to be 1.40. This deviation from quadratic behaviour may be attributed to the presence of multiple carrier scattering, such as in a semimetal. This semimetallic nature is further supported to the first-principle calculations, which we discussed in the section-III.  

Next, we investigated the magnetotransport as a function of applied field across the transition at different constant temperatures. 
Furthermore, \hyperref[fig4]{Fig. 7(b)} displays longitudinal magnetoresistance with different magnetic fields up to 8 T.  With a maximum applied magnetic field of 8 T, a positive non-saturating magneto-resistance behaviour is seen at 2 K, with an MR of $\approx$ 2\%. Usually, positive MR in metals observed due to the dominance of orbital contribution. 
Below the transition temperature ($T_N$) the MR decreases drastically, which may be ascribed to the AFM ordering in the GdGaSi. The obtained MR\% is relatively lower than the previously reported polycrystalline sample of GdAlGe (23\%) \cite{WANG2021526}.  

According to the classical MR model, magnetoresistance curves display quadratic field dependence i.e. MR $\propto H^2$  for normal metals and semiconductors \cite{Rossiter2006}. 
In our case, we have observed that MR data have shown distinct field dependence, having drastic decrease in MR magnitude across the magnetic transition.
To investigate the field dependence in magnetoresistance curves, we employed Kohler's scaling analysis. 
The phonon-scattering induced damping in the MR magnitude can be normalized through the Kohler's scaling, which scales applied field with zero field resistivity at the temperature of measurement ($H/\rho_0$).
Thus, Kohler’s rule implies that if the scattering mechanism across the Fermi surface remains constant throughout the temperature region, then MR vs $H/\rho_0$, for various temperatures should fall onto one single curve. 

Furthermore, the obtained Kohler's plots exhibit separation into two branches for curves, (1) 2 K $<T<T_N$ and (2) 300 K$>T>T_N$. This feature of separation across transition is consistent with observation of MR as function of temperature (see inset (I) of \hyperref[fig5]{Fig. 8(a)}).
It is describable from the \hyperref[fig5]{Fig. 8(a)} that Kohler plots for $T<T_N$ fall on one single curve while for all curves at higher temperatures there is a different power dependence which is consistent up to 300 K.  This may be due to the magnetic ordering which affects the scattering time of the carrier \cite{Singh2024}. In an attempt to fit the MR data as function of field, we have utilized the following relation;

\begin{equation}
MR = \alpha \left( \left( \frac{H}{\rho_0} \right)^m \right)
\end{equation}  

As shown in the \hyperref[fig5]{Fig. 8(b,c)}, we selected two representative temperatures on either side of the transition to illustrate the field dependence. Notably, the exponent $m$ changes from 1.48 at 5 K to 2 at 25 K. The quadratic field dependence observed at higher temperature ($>T_N$) aligns well with the expected metallic behaviour and trivial charge carriers. More intriguingly, the deviation from this quadratic trend at low temperatures ($<T_N$) suggests the presence of non-trivial electronic characteristics. This observation strengthens the possibility of non-trivial characteristics such as linear dispersion in this class of material, which has been established earlier in other isostructural analogs of this series GdAlSi \cite{Laha2024109}.

\subsection{Hall resistivity studies}

Motivated by distinct charge carrier dynamics across magnetic transition supported from Kohler's analysis of MR data, we performed Hall measurements at various temperatures across transition in the applied field range of $\pm$ 8 T. 
The Hall data have been antisymmetrized to remove any MR contribution, through $\rho_{xy} = [\rho_{xy}(+H)-\rho_{xy}(-H)]/2$, for different temperatures. 
Hall resistivity exhibits a linear behaviour throughout the field range signifies the presence of one type of charge carrier. In the inset of \hyperref[fig6]{Fig. 9(a)}, $\rho_{xy}$ vs $H$ for the temperature T = 2 K has been fitted by $\rho_{xy}=R_0H$, where $R_0$ is the Hall coefficient.
The negative value of $R_0$ signifies that electron is the majority charge carrier in GdGaSi. 
Again, carrier concentration and mobility can be calculated using the formula $n_e=\frac{1}{R_0e}$ and $\sigma=n_ee\mu_e$, respectively. Where $n_e$ is the carrier concentration, $\mu_e$ is the mobility of electrons and $\sigma=1/\rho$ is the longitudinal conductivity. The semi-metallic behaviour of GdGaSi is indicated by the value of $n_e$ ($\approx 10^{21}$),  which is less than that of ordinary metals ($10^{22}-10^{23}$).  However, the value of $n$ is equivalent to the ZrSiS type nodal line semimetals \cite{Hu2016}, but much bigger than other Dirac semimetals like Cd$_3$As$_2$ ($10^{16}-10^{17}$) and Na$_3$Bi ($10^{17}-10^{18}$) \cite{Feng2017, Xiong2015, He2014}. The observed semimetallic nature supports our theoretical findings. 

Further, the values of $n$ and $\mu$ of GdGaSi are consistent with other members of this family \cite{Yang2021,Lyu2020}. 
As shown in \hyperref[fig6]{Fig. 9(b,c)}, above magnetic ordering temperature $T_N$, the carrier density fluctuates and have small dependence as function of temperature up to 300 K, while the Hall mobility decreases  monotonously with increasing temperature due to increased carrier scattering at higher temperature. 
Below $T_N$, in the magnetic ordered state, the scattering is suppressed gradually with lowering the temperature, leading to increased mobilities at lower temeprature. The charge concentration follows an opposite trend and decreases below magnetic transition, it may be an effect of modulation in the Fermi surface topology in the ordered state. 
With the available experimental results, it remains inconclusive whether the observed trend in charge carrier mobility is solely due to reduced scattering or if it also involves the presence of highly mobile fermions distinct from the carriers dominant at higher temperatures.
These observations signal the possibility of GdGaSi as a Weyl semimetal with non-trivial characteristics as its other isostructural compounds.



\section{CONCLUSIONS}
In summary, first-principles calculations were performed to explore the magnetic ground state and electronic structure of the material. The ground state was found to be the AFM1 configuration, in which the Gd spins exhibit intraplane ferromagnetic and interplane antiferromagnetic coupling. Furthermore, the band structure calculations reveal the semimetallic nature of the system, with the presence of both hole and electron pockets near the Fermi level. Notably, a dominance of electron-like carriers is observed for the compound. The presence of interesting band crossings near the Fermi level suggests the potential for non-trivial topological features. 
To validate the theoretical result, we have prepared the GdGaSi compound and investigated its physical properties experimentally. DC magnetization study reveals the AFM nature below 19 K, and having signature of spin-reorientation below $T_N$. 
Semimetallic nature of the sample is confirmed from the longitudinal resistivity data having drop below 19 K consistent with magnetic ordering from $M(T)$ data. Linear and negative field dependence of Hall resistivity support the electrons as the majority charge carrier. The impact of magnetic ordering on the scattering mechanism is evident from the Kohler's analysis and change in mobility of carriers, which supports our theoretical findings.
On the basis of observed characteristics, GdGaSi is candidate material to be explored as single crystal and monolayer for understanding correlation between broken inversion symmetry and non-triviality in ordered magnetic state.


\section*{ACKNOWLEDGMENTS}
M.P. would like to acknowledge NIT Andhra Pradesh for the fellowship. T.P. would like to acknowledge UGC DAE CSR (Grant No. CRS/2021-22/02/487) and SERB (Grant No. CRG/2022/008197) for their financial support. SM acknowledges DST, India. S.P. and V.K. sincerely acknowledge the National Supercomputing Mission (NSM) for providing computing resources of ‘PARAM SEVA’ at IIT, Hyderabad. S.P. acknowledges  DST-INSPIRE for a research fellowship. The authors thank Dr. S. Isogami at NIMS in Japan for his technical support.

%

\end{document}